\documentclass[seceq,supplement]{ptptex}

\title{
Gravitational radiation reaction%
}

\author{
Takahiro \textsc{Tanaka} 
}

\inst{
Department of Physics,~Graduate School of Science,
Kyoto University,\\ Kyoto 606-8502,~Japan
}

\abst{
We give a short personally-biased review on the recent progress 
in our understanding of gravitational radiation reaction acting 
on a point particle orbiting a black hole. 
The main motivation of this study is to obtain sufficiently 
precise gravitational waveforms from inspiraling 
binary compact stars with a large mass ratio. 
For this purpose, various 
new concepts and techniques have been developed 
to compute the orbital evolution 
taking into account the gravitational self-force. 
Combining these ideas with a few supplementary new 
ideas, we try to outline a path to our goal here. 
}

\begin{document}

\maketitle

\section{Introduction}
Ground-based interferometric gravitational wave detectors have 
started operation and are also in the phase of 
rapid improvement~\cite{LIGO,TAMA,GEO,VIRGO}. 
R\&D studies of a
space-based gravitational wave observatory project, the Laser
Interferometer Space Antenna (LISA)~\cite{LISA}, which observes in
the {\rm mHz}-band, are in rapid progress. There is also proposal
for a DECi hertz Interferometer Gravitational wave Observatory
(DECIGO/BBO)~\cite{DECIGO,BBO}, which will be a laser interferometer
gravitational wave antenna in space sensitive at 
$ f\sim 0.1{\rm Hz}$.

Among the promising targets for space interferometers, 
the most precise theoretical prediction and observational 
measurement of gravitational waveform are expected from
the inspiral stage 
of binary systems comprising of a supermassive black hole 
($M \sim 10^{5-8} M_{\odot}$) and a compact
object of solar mass ($\mu \sim 1-10 M_{\odot}$)\cite{Cutler}. 
Here the meaning of ``precise'' is two-fold. 
One is that the number of cycles 
is large and the other is that higher order post-Newtonian 
corrections are necessary. 
These gravitational wave sources can provide the first
high-precision test of general relativity in very strong
gravitating regimes\cite{test}.

On the theoretical side, to obtain sufficiently precise templates, 
the standard post-Newtonian approximation\cite{PN-latestreview} 
seems to be insufficient since the accessible highest order of expansion 
is practically limited. 
However, there is a natural expansion parameter in this system, 
that is the mass ratio $\mu/M$.
Therefore, in place of the standard post-Newtonian expansion, 
there is a possibility to develop perturbation theory 
based on the well-understood black hole linear 
perturbation theory
\cite{RW,Zerilli,Teukolsky,SasNak,Chandra,Sasaki-Tagoshi}
 for binaries with extreme mass ratios. 
To develop a method to 
compute waveforms on this line is our
purpose of studying gravitational radiation reaction. 

This paper is organized as follows. 
In Sec.~\ref{energybalance}, we briefly discuss the very basics of 
black hole perturbation approach, 
followed by a discussion on balance 
argument about radiation reaction, given in Sec. 3. 
In Secs.~4 and 5, we consider radiation reaction in 
adiabatic approximation. The basic idea is explained in 
Sec.~4, 
while the recent developments, which have led to 
a very concise formula for the change rate of Carter 
constant, are presented in Sec.~5. 
A method for long time integration is also discussed. 
In Secs.~6 and 7, we discuss the instantaneous self-force. 
The main stream of the procedure about how to evaluate the self-force, 
although biased by my personal view, is explained in Sec.~6, 
supplemented with comments on some topics given in Sec.~7. 
The discussions presented in Sec.~5 might be beyond 
the level of an introductory review paper. 
Section 8 is devoted to Summary. 

In this paper we use the units in which $8\pi G_N=1$. 

\section{Before radiation reaction}
\label{energybalance}
We model a binary system by a point particle 
of mass $\mu$ orbiting a black hole of mass $M$
assuming $\mu\ll M$. 
In the lowest order in the mass ratio, $O((\mu/M)^0)$, 
the particle moves along a geodesic on the background geometry. 
Using the black hole perturbation, we can compute the energy 
and angular momentum flux carried by gravitation waves emitted 
by the particle to the infinity or into the horizon.  
Already in this lowest order approximation,
this black hole perturbation approach has proven 
to be very powerful for evaluating 
general relativistic gravitational waveforms. 

The perturbed metric is expanded as 
\begin{eqnarray*}
 g_{\mu\nu}=g_{\mu\nu}^{BH}+h_{\mu\nu}^{(1)}+h_{\mu\nu}^{(2)}+
      \cdots.
\end{eqnarray*}
Then the linearized perturbed Einstein equations 
\begin{eqnarray*}
\delta G_{\mu\nu}[{\bf h}^{(1)}]=T^{(1)}_{\mu\nu},
\end{eqnarray*}
become coupled equations for metric perturbations ${\bf h}^{(1)}$ 
in general, and are difficult to solve. 
However, when the background spacetime is 
given by Schwarzschild or Kerr black hole, 
perturbation equations can be 
written in terms of a single master equation due to symmetry. 
Here we schematically write the master equation as
\begin{eqnarray*}
L\zeta^{(1)}=\sqrt{-g}\, T^{(1)},  
\end{eqnarray*}
where $L$ is a second order differential operator and 
$\zeta^{(1)}$ is a master variable in the linear order. 
The source $T^{(1)}$ is a function 
obtained from the energy momentum tensor. 

This reduction is called 
Regge-Wheeler-Zerilli formalism 
in the Schwarzschild case\cite{RW,Zerilli},
and Teukolsky-Sasaki-Nakamura 
formalism in the Kerr case\cite{Teukolsky, SasNak}. 
In the latter case, starting with metric perturbations is 
not very successful. Instead, we use Newman-Penrose quantities 
defined by, say, 
\begin{equation}
 {}_{-2}\psi\approx - C_{\mu\nu\rho\sigma}
   n^\mu \bar m^\nu n^\rho \bar m^{\sigma}, 
\end{equation}
as a master variable. This variable is a contraction of 
Weyl tensor $C_{\mu\nu\rho\sigma}$ with two null 
tetrad bases, $n^\mu$ (in-going) and $\bar m^\mu$ (angular). 
Here note that $m^\mu$ is complex valued and 
``$~\bar~~$'' represents complex conjugation. 
We can similarly define ${}_2\psi$ by considering the 
contraction with $\ell^\mu$ (out-going) and $m^\mu$. 
As a result, we have two equations 
\begin{equation}
 {}_s L \,_s\psi=\sqrt{-g}\,_sT, 
\qquad (s=\pm 2)
\end{equation}
but they are not mutually independent. 
The source ${}_sT$ is obtained by applying 
a corresponding second order  
differential operator ${}_s\tau_{\mu\nu}$
to the energy momentum tensor $T^{\mu\nu}$. 

Thus obtained master equation is separable. 
We introduce a harmonic function which describes 
angular dependence $S_{\Lambda}(\theta,\varphi)$, 
where $\Lambda=\{\ell,m,\omega\}$ is a set of two angular 
eigenvalues and one frequency eigenvalue. 
Homogeneous solution can be found in the form of a linear 
combination of mode functions 
${}_s\Omega_\Lambda=
{}_sR_{\Lambda}(r)
 \,{}_s S_{\Lambda}(\theta,\varphi)\,
  e^{-i\omega t}
$. Substituting this form, the problem to find a 
homogeneous solution reduces to solving a second order 
ordinary differential equation
\begin{equation}
 \left[\partial_r^2+\cdots\right]{}_sR_\Lambda(r)=0.
\end{equation}

By using homogeneous solutions which satisfy 
appropriate boundary conditions, we can construct 
Green function. Then, the solution with source 
is given by 
\begin{equation}
 {}_s\psi\approx 
 \sum_\Lambda{1\over {}_sW_{\Lambda}}
{}_s\Omega^{up}_{\Lambda}(x)
\int \sqrt{-g} d^4x' {}_s\bar{\tilde\Omega}^{out}_{\Lambda}(x')
   T(x')\theta(r-r')+\cdots,
\end{equation}
where {\it up} and {\it out}, respectively, 
mean that there is no in-coming wave from 
the past infinity and that there is no wave absorbed 
into the future horizon.
Here, ``$\cdots$'' stands for the part which contains $\theta(r'-r)$.  
We have introduced spin-flipped mode functions,  
${}_s\tilde\Omega_{\Lambda}(x')={}_{-s} R_{\Lambda}(r)
 \,{}_s S_{\Lambda}(\theta,\varphi)\, e^{-i\omega t}$.  
Notice that 
${}_s\bar {\tilde\Omega}^{out}_{\Lambda}(x')
 ={}_{s} R^{in}_{\Lambda}(r)\overline{
 \,{}_s S_{\Lambda}(\theta,\varphi)\,
  e^{-i\omega t}}$. (Since the differential operator $L$
for the Teukolsky equation 
is not real, a simple complex conjugation does not give a 
solution of the same equation but it becomes a solution 
of the equation with its spin flipped from $s$ to $-s$.
Therefore we cannot 
choose ${}_s R^{in}={}_{s}\bar R^{out}$ as usual, but we can choose 
${}_s R^{in}={}_{-s}\bar R^{out}$. )
${}_sW_{\Lambda}$ is the Wronskian 
between these two radial functions, i.e., 
\begin{equation}
 {}_sW_{\Lambda}=W({}_s R^{up}_\Lambda,{}_s R^{in}_{\Lambda}(r))
    \approx 
  {}_sR^{up}_\Lambda(r){\partial\, 
            {}_s R^{in}_{\Lambda}(r)\over\partial r}
  -{}_s R^{in}_\Lambda(r){\partial\, {}_sR^{up}_{\Lambda}(r)
               \over\partial r}.
\label{Wronskian}
\end{equation}
There is a systematic method to 
solve homogeneous Teukolsky equation as a series 
expansion~\cite{ManoTak,ManoTak2,Sasaki-Tagoshi}. 
Using this method, one can rather easily obtain the solution, and 
we can write down the expression analytically in explicit form 
once we invoke the post-Newtonian expansion. 

To estimate the fluxes at $r\to \infty$, one can use a simple relation, 
$_{-2}\psi\approx {1\over 2}(\ddot h_+-i\ddot h_\times)$. 
Using this relation, 
one can estimate, say, the energy loss rate due to 
gravitational wave emission per unit steradian as 
\begin{eqnarray*}
 \mu {d^2 {\cal E}_{GW}\over dtd\Omega} 
 ={r^2\over 4\pi \omega^2}|_{-2}\psi|^2. 
\end{eqnarray*}
In this paper, we use ${\cal E}$ (${\cal L}$) as the specific energy 
(angular momentum) per unit mass of the particle. 
The contribution from the waves absorbed by the black hole 
can be evaluated in a similar manner. 

\section{Balance argument and its limitation}
When radiation reaction is not very significant, 
the trajectory of a particle is almost a geodesic. 
In such a case one can estimate the evolution of orbital 
frequency by assuming that the orbit evolves 
losing its energy and angular momentum as much as those 
emitted as gravitational waves~\cite{supple}. 

In the case of circular orbits, the evolution of the 
orbital frequency almost determines the gravitational waveform 
(except for the change of amplitude). 
Equating the energy lost through the gravitational wave emission 
with the minus of the binding energy, 
the change rate of the orbital frequency $f$ is evaluated as
\begin{equation}
 {df\over dt}=-{d{\cal E}_{GW}\over dt}
\left({d{\cal E}_{orbit}\over df}\right)^{-1}. 
\end{equation}
Then the leading order of ${d{\cal E}_{GW}/ dt}$ starts with 
$O(\mu)$. (Recall that ${\cal E}$ is the specific energy. 
In counting of the order of magnitude, 
we are assuming that $M$ is $O(1)$.) 
This is because the energy was exactly conserved, 
if the radiation reaction were turned off. 
On the other hand, the leading order of the relation between 
the binding energy and the orbital frequency, 
${d{\cal E}_{orbit}/ df}$, 
is determined by the geodesic motion 
on a given background black hole spacetime. The effect of 
the self-force, which is the force acting on a particle 
due to the metric perturbation caused by the particle itself, 
is higher order correction of $O(\mu)$ in 
${d{\cal E}_{orbit}/ df}$.   
Roughly speaking, to obtain the leading order correction to the 
waveform, we therefore have only to know the effect of the 
self-force on ${d{\cal E}_{GW}/ dt}$. 
(We will return to this point in Sec.~\ref{longperiod}.)

The extension to more general orbits is straight forward
in the case of Schwarzschild background. In this case 
one can assume that the orbit is on the equatorial plane 
without loss of generality. Then, the geodesics are 
specified by the energy ${\cal E}$ and the $z$-component
of the angular momentum ${\cal L}$. Both ``constants of motion'' 
have the associated timelike and rotational Killing vectors, 
$\eta^\mu_{(t)}\equiv (\partial_t)^\mu$ and 
$\eta^\mu_{(\varphi)}\equiv(\partial_\varphi)^\mu$. 
Therefore we can define conserved current from the 
effective energy momentum tensor
$t_{\mu\nu}$, which satisfies the conservation law 
$t^{\mu\nu}{}_{\! ;\nu}=0$ with respect to the 
background covariant derivative, as 
\begin{equation}
 j^{({\cal E})}_\mu=\mu^{-1} t_{\mu\nu}\eta^\nu_{(t)}. 
\end{equation}
Then $j^{({\cal E})}_\mu$ 
satisfies $j^{({\cal E})\,;\mu}_\mu=0$, and hence one can define 
the conserved (specific) energy by the integral over 
spatial three surface $\Sigma$ as
\begin{equation}
 {\cal E}=-\int_{\Sigma} d\Sigma^\mu j_{\mu}^{({\cal E})}. 
\label{Edef}
\end{equation}
Let's choose the boundary of $\Sigma$ to be 
a sphere $S$ at a fixed radius, which is supposed 
to be sufficiently large. 
In this case the 
change rate of ${\cal E}$ is given by 
\begin{equation}
 {d{\cal E}\over dt}=-\int_S dS\, j^{({\cal E})}_{r}, 
\label{dEdt}
\end{equation}
where $dS$ is an infinitesimal element of area on $S$.  
In the above discussion, we assumed the existence of the
effective energy momentum tensor $t_{\mu\nu}$ 
that satisfies the conservation law. Now we  
directly derive it from the Einstein equation. 
The Einstein tensor can be expanded as 
\begin{eqnarray}
  G_{\mu\nu}[{\bf g}+{\bf h}] = G_{\mu\nu}^{[0]}
    +G_{\mu\nu}^{[1]}[{\bf h}]+G_{\mu\nu}^{[2]}[{\bf h},{\bf h}]
   +\cdots. 
\label{Gexpand}
\end{eqnarray}
Since the background metric ${\bf g}$ satisfies the vacuum Einstein 
equations in the present case, $G_{\mu\nu}^{[0]}=0$. 
Using this relation, the contracted 
Bianchi identity at the linear order in ${\bf h}$ becomes 
\begin{equation} 
G_{\mu\nu}^{[1]}[{\bf h}]^{;\nu}=0
\end{equation}
Here a semicolon denotes a covariant differentiation 
with respect to the {\it background} metric. 
Substituting ``${\bf h}={\bf h}^{(1)}+{\bf h}^{(2)}+\cdots$'' 
into the Einstein equations and keeping the terms up to 
second order, we obtain 
\begin{equation}
 G_{\mu\nu}^{[1]}[{\bf h}^{(1)}]+
 G_{\mu\nu}^{[1]}[{\bf h}^{(2)}]+
  G_{\mu\nu}^{[2]}[{\bf h}^{(1)},{\bf h}^{(1)}]
 = T_{\mu\nu}. 
\end{equation}
From the background covariant derivative of this equation, 
we have 
\begin{equation}
  G_{\mu\nu}^{[2]}[{\bf h}^{(1)},{\bf h}^{(1)}]^{;\nu}
 =T_{\mu\nu}^{~~;\nu}. 
\end{equation}
Hence we find that 
\begin{equation}
t_{\mu\nu}\equiv 
 T_{\mu\nu}-G_{\mu\nu}^{[2]}[{\bf h}^{(1)},{\bf h}^{(1)}]
\end{equation}
satisfies the conservation law with respect to the background 
covariant derivative. One may identify the second term as 
the effective energy momentum tensor of gravitational waves
$t_{\mu\nu}^{(GW)}$. Accordingly, one can divide the energy ${\cal E}$ 
into two parts: ${\cal E}={\cal E}_{orbit}+{\cal E}_{GW}$. ${\cal E}_{orbit}$ is the contribution 
from $T_{\mu\nu}$ while ${\cal E}_{GW}$ is that from $t_{\mu\nu}^{(GW)}$. 

Now we should note that 
${\cal E}_{orbit}$ is $O(1)$ while 
${\cal E}_{GW}$ is $O(\mu)$.
Keeping this fact in mind, 
we look at Eqs.~(\ref{Edef}) and (\ref{dEdt}) again. 
Then ${\cal E}$ is $O(1)$ and the leading term comes 
from ${\cal E}_{orbit}$. 
On the other hand, $d{\cal E}/dt$ does not have contribution 
from $T_{\mu\nu}$, and hence it is $O(\mu)$. 
However, we do not conclude that the radiation reaction 
is unimportant. 
After integration over a long period of $O(\mu^{-1})$
the change in ${\cal E}$ becomes $O(1)$. This change in ${\cal E}$ 
must be attributed to the change of $T_{\mu\nu}$ because 
${\cal E}_{GW}$ should stay $O(\mu)$.  
This consideration establishes the balance argument to 
the motion of a particle:
\begin{eqnarray}
 {d{\cal E}_{orbit}\over dt}\approx -\int_S dS\, j_{r}^{({\cal E})}. 
\label{dEorbit}
\end{eqnarray}

In the case of Kerr background, we do not have 
spherical symmetry. Hence, one cannot say that 
the orbits are on the equatorial plane in general. 
To specify geodesics off the equatorial plane, we need 
to consider another ``constant of motion''. 
It is well known that the geodesics on the Kerr background 
possess the third ``constant of motion'' called Carter 
constant. However, Carter constant is not associated with 
any Killing vector field. Instead, it is related to 
a rank two Killing tensor.  

Let us examine the ``constants of motion'' in Kerr in more detail. 
The background Kerr spacetime in the 
Boyer-Lindquist coordinates is given by 
\begin{eqnarray}
ds^2& = & -\left(1-{2Mr\over\Sigma}\right)dt^2
-{4Mar\sin^2\theta\over \Sigma}dtd\varphi\cr
&& +{\Sigma\over \Delta}dr^2+\Sigma d\theta^2
+\left(r^2+a^2+{2Ma^2r\sin^2\theta\over \Sigma}\right)
\sin^2\theta d\varphi^2, 
\end{eqnarray}
where Kerr parameter $a$ is the the angular momentum of
the black hole divided by its mass, 
$
\Sigma:=r^2+a^2\cos^2\theta
$, and $
\Delta:=r^2-2Mr+a^2.
$
Killing tensor in Kerr spacetime is given by 
\begin{equation}
K_{\mu\nu}:
=2\Sigma l_{(\mu}n_{\nu)}+r^2g_{\mu\nu},
\end{equation}
where the parentheses
denote symmetrization on the indices enclosed, and
$
l^{\mu}:=
\left({r^2+a^2},\Delta,0,a \right)/\Delta
$
and
$
n^{\mu}:=
\left(r^2+a^2,-\Delta,0,a\right)/2\Sigma
$
are out-going and in-going radial null vectors, respectively.
Killing tensor satisfies the equation 
\begin{equation}
 K_{(\mu\nu;\rho)}=0.
\label{Ksym}
\end{equation}
Using this Killing tensor, the Carter constant is defined as 
\begin{equation}
{\cal Q}\equiv K_{\alpha\beta}u^{\alpha}u^{\beta},
\label{eq:Carter}
\end{equation}
where $u^{\alpha}:=dz^{\alpha}/d\tau$ is the four velocity 
of an orbiting particle.
We often use another notation for the Carter constant, 
${\cal C} :=  {\cal Q}-(a{\cal E}-{\cal L})^2$, defined 
in such a way that it vanishes for orbits on the equatorial plane. 
By using the symmetry of the Killing tensor, it is 
easy to check that thus defined Carter constant 
does not vary along geodesic. For Carter constant, 
however, we cannot define a quantity corresponding to 
${\cal E}_{GW}$. Therefore there is no counter part of 
Eq.~(\ref{dEorbit}) for $d{\cal Q}/dt$.  

\section{Adiabatic approximation for $d{\cal Q}/dt$}

As we have seen in the previous section, 
one cannot use the balance argument 
to evaluate the change rate of ${\cal Q}$. 
Then, to evaluate $d{\cal Q}/dt$,
we need to compute the self-force acting on the particle 
directly~\cite{Ori95}.
Though the prescription to calculate the self-force is formally
established~\cite{DB,MST97,QW97}, performing 
explicit calculation is not so straight forward. 
However, it was found to be easier to compute the 
averaged value of $d{\cal Q}/dt$. As an approximation, we may 
use the averaged values of the change rates for the ``constants 
of motion'' instead of those evaluated by using the instantaneous 
self-force. We call it adiabatic approximation. 
This approximation will be as good as an estimate using 
the balance argument, and moreover 
is also applicable to $d{\cal Q}/dt$. 

\subsection{use of radiative field}
Gal'tsov~\cite{Gal'tsov82} advocated to use 
the \emph{radiative} part of the metric perturbation, 
which was introduced earlier by 
Dirac\cite{Dirac38}, 
to calculate $d{\cal E}/dt$ and $d{\cal L}/dt$. 
The radiative field is defined by half retarded field 
minus half advanced one. The divergent part contained 
in the retarded field is common to that contained in the 
advanced field. Therefore the combination of the 
radiative field is free from divergence. Hence as far as 
we discuss the self-force composed of the radiative field, 
we do not have to worry about how to regularize the force. 
The question however remains whether the radiative self-force correctly 
reproduces the result which is obtained by using the retarded field 
with the aid of appropriate regularization procedure. 

Gal'tsov has shown that the radiative field correctly 
reproduces the results 
obtained by using the balance argument for 
$d{\cal E}/dt$ and $d{\cal L}/dt$ 
when they are averaged over an infinitely long time interval 
assuming a geodesic motion as a source of metric perturbation. 
When $\mu$ is thought to be an infinitesimal expansion parameter, 
the trajectory of the particle follows exactly a 
geodesic at the lowest order. In this sense, in the lowest order, 
this averaging over a long period of time 
assuming a geodesic motion is justified 
when we evaluate the self-force to the lowest order in $\mu$. 

However, there had been no justification for  
applying the same scheme to $d{\cal Q}/dt$ until very recently.  
The breakthrough was brought by Mino, 
who gave a justification for applying the same scheme
to $d{\cal Q}/dt$~\cite{Mino03}. (See also Ref.~\citen{Hughes05}).
Namely, he has proven that
\begin{equation}
 \left\langle {d\over d\tau} {\cal Q}\right\rangle
 ={1\over \mu}\lim_{T\to\infty}
 {1\over 2T}\int_{-T}^T d\tau {\partial {\cal Q}\over \partial u^\alpha}
   F^\alpha\left[{\bf h}^{(rad)}\right], 
\label{dQdtau}
\end{equation}
where $F^\alpha\left[{\bf h}^{(rad)}\right]$ is 
the self-force evaluated by using the radiative field 
${\bf h}^{(rad)}\equiv ({\bf h}^{(ret)}-{\bf h}^{(adv)})/2$. 

The key observation in his proof is the invariance of 
the geodesics under the transformation:
\begin{eqnarray*}
 a\to -a, \qquad
 (t,r,\theta,\varphi)\to (-t,r,\theta,-\varphi).
\end{eqnarray*}
All geodesics, unless the 
values of $({\cal E,L,Q})$ are fine tuned, 
transform into themselves under the above transformation 
if we choose the origin of $t$ and $\varphi$ appropriately. 
Once we notice this symmetry, it is easy to show that 
\begin{equation}
\left\langle \left({d{\cal Q}\over d\tau}\right)^{(ret)}\right\rangle
=-\left\langle \left({d{\cal Q}\over d\tau}\right)^{(adv)}\right\rangle, 
\label{retadv}
\end{equation}
where the superscripts $(ret)$ and $(adv)$ indicate
that the retarded field ${\bf h}^{(ret)}$ 
and the advanced field ${\bf h}^{(adv)}$ are used 
instead of ${\bf h}^{(rad)}$ in Eq.~(\ref{dQdtau}). 
The relation (\ref{retadv}) justifies the use of 
the formula~(\ref{dQdtau}). 

The radiative Green function has a simple structure which 
does not contain any step function $\theta(r-r')$:
\begin{eqnarray}
G^{(rad)}(x,x')&=& \sum_{\Lambda}{1\over
    W({}_s R^{up}_\Lambda,{}_s R^{in}_\Lambda)
    W({}_s R^{down}_\Lambda,{}_s R^{out}_\Lambda) }
\cr &&\hspace{-2cm}\times
    \left(
    W({}_s R^{in}_\Lambda,{}_s R^{out}_\Lambda) 
     {}_s\Omega^{down}_{\Lambda}(x) 
     {}_s\bar{\tilde\Omega}^{down}_{\Lambda}(x')
   +W({}_s R^{down}_\Lambda,{}_s R^{up}_\Lambda) 
    {}_s\Omega^{out}_{\Lambda}(x) 
    {}_s\bar{\tilde\Omega}^{out}_{\Lambda}(x'). 
    \right),  \cr
&& \label{eq2}
\end{eqnarray}
To show this, let us start with the following defining expression
of the radiative Green function for $r>r'$;
\begin{equation}
 \sum_{\Lambda}e^{-i\omega(t-t')}
   \left[
 \frac{{}_s R^{up}_\Lambda(r){}_s R^{in}_\Lambda(r')}
      {W({}_s R^{up}_\Lambda,{}_s R^{in}_\Lambda)}-
 \frac{{}_s R^{down}_\Lambda(r){}_s R^{out}_\Lambda(r')}
    {W({}_s R^{down}_\Lambda,{}_s R^{out}_\Lambda)}
 \right] 
    {}_s S_{\Lambda}(\theta,\varphi) 
   {}_s \bar S_{\Lambda}(\theta',\varphi'). 
\label{eq1}
\end{equation}
Since our goal is to obtain an expression in terms of 
$down$-field and $out$-field, we want to 
eliminate ${}_s R^{up}_\Lambda(r)$
and ${}_s R^{out}_\Lambda(r')(={}_{-s} \bar R^{in}_\Lambda(r'))$ 
in Eq.~(\ref{eq1}). Hence we expand 
${}_s R^{up}$ and ${}_{s}R^{out}$ as  
\begin{eqnarray}
 {}_s R^{up} 
 & = & \alpha~ {}_s R^{out}+\beta~ {}_s R^{down}\cr
 {}_s R^{out} & = & \gamma~ {}_s R^{up}+\delta~ {}_s R^{in}. 
\label{expansion}
\end{eqnarray}
Taking the Wronskian of both sides of Eqs.~(\ref{expansion})
with appropriate radial functions, 
one can easily obtain 
\begin{eqnarray*}
 && W({}_s R^{up},{}_s R^{down})
   =\alpha\, W({}_s R^{out},{}_s R^{down}),
\qquad
  W({}_s R^{up},{}_s R^{out})
   =\beta\, W({}_s R^{down},{}_s R^{out}), 
\cr
  &&   W({}_s R^{out},{}_s R^{in})
   =\gamma\, W({}_s R^{up},{}_s R^{in}),
\qquad
  W({}_s R^{out},{}_s R^{up})
   =\delta\, W({}_s R^{in},{}_s R^{up}). 
\end{eqnarray*}
Substituting these relations, the expression (\ref{eq1}) 
reduces to (\ref{eq2}). 
We can do an analogous reduction for $r<r'$, and the result 
turns out to be the same as that 
for $r>r'$. Namely, the step functions 
which was present in the retarded and the advanced Green functions 
do not appear in the radiative Green function. 
This means that the radiative field is a source-free 
homogeneous solution. 

\subsection{metric reconstruction\cite{sago}}
\label{metricreconst}
In order to use the formula~(\ref{dQdtau}), we need to know 
how to reproduce the metric perturbations from 
the master variable ${}_s\psi$. 
In the present case, what we have to deal with 
is a source-free homogeneous 
solution. This fact simplifies the reconstruction significantly
\footnote{ 
When we reconstruct the metric perturbation 
from a solution of the master variable within the source distribution, 
more delicate treatment is necessary~\cite{Ori02}.
Below we use the assumption of the factorized form of 
the tensor Green function. 
This assumption is not necessary 
if we follow the derivation given by Wald\cite{Wald}, and 
actually this assumption itself is not correct\cite{Ori02}.  
}.
The method was originally given by Chrzanowski~\cite{Chr,Wald}. 
Here, we present the basic idea of the derivation of 
reconstruction formula, neglecting details. 

We formally write the metric perturbation induced by 
a source $T^{\alpha\beta}$ as 
\begin{equation}
 h_{\mu\nu}(x)
  =\int\sqrt{-g}d^4 x'\,
  G_{\mu\nu\alpha\beta}^{(ret)}(x,x') T^{\alpha\beta}(x'). 
\end{equation}
We assume that $T^{\alpha\beta}(x')$ is localized within $r'<r_0$.  
And we assume a factorized form of the tensor Green function as 
\begin{equation}
 G_{\mu\nu\alpha\beta}^{(ret)}(x,x')
 =\sum_\Lambda {1\over {}_s{\cal N}\, {}_sW_\Lambda}
   \left(\Pi^{up}_{\Lambda\mu\nu}(x)
         \bar\Pi^{out}_{\Lambda\alpha\beta}(x')
   \theta(r-r')+\cdots\right), 
\end{equation}
where $\Pi_{\mu\nu}$ is the mode function for the 
metric perturbation, whose explicit form is unknown 
at this point. Since we do not know how to normalize the 
mode function for metric perturbation, we have introduced 
a constant ${}_s{\cal N}$ to take care of this normalization. ${}_sW_\Lambda$ 
is the Wronskian for the corresponding mode function for the 
master variable defined by (\ref{Wronskian}). 

The master variable can be computed from the metric perturbation 
by applying a second order differential operator as 
${}_s\Omega_{\Lambda}={}_s D^{\mu\nu}\Pi_{\Lambda\mu\nu}$. 
Hence, we have 
\begin{equation}
 {}_s\psi={}_s D^{\mu\nu} h_{\mu\nu}
   =\sum_{\Lambda} {1\over {}_s{\cal N} {}_sW_\Lambda}
    {}_s\Omega^{up}_{\Lambda}(x)
\int\sqrt{-g}\,d^4x'\,
       \bar\Pi^{out}_{\Lambda\alpha\beta}(x')T^{\alpha\beta}(x'), 
\end{equation}
for $r>r_0$, outside the source distribution.
On the other hand, we can evaluate the same quantity 
by using the Green function for the master variable, which 
leads to 
\begin{equation}
{}_s\psi={}_sD^{\mu\nu} h_{\mu\nu}(x)
 =\sum_\Lambda {1\over {}_sW_\Lambda}
   {}_s\Omega^{up}_{\Lambda}(x)
   \int\sqrt{-g}\, d^4x'\,
       {}_s\bar{\tilde\Omega}^{out}_{\Lambda}(x'){}_sT(x'). 
\end{equation}
Comparing these two expressions, we find 
\begin{equation}
\int\sqrt{-g}\, d^4x'\,
       \Pi_{\Lambda\mu\nu}^{out}(x')T^{\mu\nu}(x')
 = {}_s\bar{\cal N} \int\sqrt{-g}\, d^4x'\,
       {}_s{\tilde\Omega}^{out}_{\Lambda}(x')
       {}_s\bar\tau_{\mu\nu} T^{\mu\nu}(x'). 
\end{equation}
The same relation holds for the modes with other boundary conditions. 
Since this relation holds for arbitrary $T_{\mu\nu}$ as long as 
it satisfies the conservation law, $T^{\mu\nu}_{~~~;\nu}=0$, 
one can establish the relation 
\begin{equation}
\Pi_{\Lambda\mu\nu}(x)
 = {}_s\bar{\cal N} \,{}_s\tau_{\mu\nu}^*\,
     {}_s\tilde \Omega_{\Lambda}(x'),
\label{OmegaPi}
\end{equation}
where ${}_s\tau^*_{\mu\nu}$ 
is the second order differential 
operator that is obtained by integration by parts from 
${}_s\bar \tau_{\mu\nu}$. 
This equation still has an ambiguity of adding pure gauge term 
$\xi_{(\mu;\nu)}$ since its contraction with $T^{\mu\nu}$ 
vanishes after integration over four volume. 

The above is a very crude explanation about the reason 
why we can reconstruct the 
metric perturbation form the master variable just by 
operating a second order differential operator. 
More rigorous derivation was given by Wald~\cite{Wald}.
With the aid of Starobinsky-Teukolsky identity\footnote{
$``{}_sD^{\mu\nu} \,{}_s\tau_{\mu\nu}^*$'' reduces to a 
forth order differential operator which transforms 
the radial function ${}_{-s} R_{\Lambda}$ to the 
spin-flipped one ${}_{s} R_{\Lambda}$. 
This identity is called Starobinsky-Teukolsky identity~\cite{Chandra}. 
}, we can 
explicitly show that
\begin{equation}
{}_s\Omega_\Lambda=
 {}_sD^{\mu\nu}{}_s\Pi_{\Lambda \mu\nu}
  = {}_sD^{\mu\nu} {}_s\bar{\cal N} \,{}_s\tau_{\mu\nu}^*\, 
   {}_s\tilde \Omega_\Lambda, 
\label{PitoOmega}
\end{equation}
simultaneously fixing the normalization constant ${}_s{\cal N}$. 

When the source is composed of a point particle, i.e., 
when the energy-momentum tensor takes the form 
$T^{\mu\nu}=\int d\tau (-g)^{-1/2} $
$u^\mu u^\nu\delta^4(x-z(\tau))$, 
the radiative metric perturbation is given by 
\begin{eqnarray}
h_{\mu\nu}^{(rad)} &= & \sum_{\Lambda}{
   1\over 
    {}_s\bar{\cal N}\, W({}_s R^{up}_\Lambda,{}_s R^{in}_{\Lambda})
    W({}_s R^{down}_\Lambda,{}_s R^{out}_\Lambda) }\cr
 && \times  \Bigl(
    W({}_s R^{in}_\Lambda,{}_s R^{out}_\Lambda) 
     {}_s\Pi^{down}_{\Lambda \mu\nu}(x) 
    \int {d\tau\over \Sigma}\, \bar \phi_{\Lambda}^{up}(z(\tau))\cr
&&\qquad\qquad
   +W({}_s R^{down}_\Lambda,{}_s R^{up}_\Lambda) 
    {}_s\Pi^{out}_{\Lambda \mu\nu}(x) 
    \int {d\tau\over \Sigma}\, \bar \phi_{\Lambda}^{in}(z(\tau))
    \Bigr)+(c.c.).
\label{hrad}
\end{eqnarray}
where
\begin{equation}
 \phi^{up}_{\Lambda}
=\Sigma \tilde u^{\mu} \tilde u^{\nu} 
    \Pi^{down}_{\Lambda\mu\nu}
={}_s\bar{\cal N}\Sigma\, \tilde u^{\mu} \tilde u^{\nu} 
    {}_s\tau^{*}_{\mu\nu}\, {}_s\bar\Omega^{up}_{\Lambda\mu\nu}, 
\label{hrad3}
\end{equation}
and $\phi^{up}$ is defined in a similar manner. 
For future convenience, instead of the four velocity $u^\mu$, 
we used $\tilde u^\mu$, an extension of $u^\mu$ to a vector field, 
whose definition is given below Eq.~(\ref{dqdt}). 
The factor $\Sigma(=r^2+a^2\cos^2\theta)$ 
is also introduced for future convenience. 

We can easily confirm that 
${}_sD^{\mu\nu} h_{\mu\nu}^{(rad)}$ with the substitution of the 
above expression reproduces 
the same ${}_s\psi^{(rad)}$ that is obtained 
by using Eq.~(\ref{eq1}) neglecting the last complex conjugate 
term. This term is necessary to make the expression real. 
To show that adding this term does not disturb the property 
of reproducing ${}_s\psi^{(rad)}$, a more detailed discussion 
is necessary\cite{Gal'tsov82}. However, we will not go into 
such a technical detail here.

\section{Simplified $d{\cal Q}/dt$ formula}

Now we know how to compute $d{\cal Q}/dt$ in principle. 
However, actual implementation of calculation 
is not so straight forward. 
In this section we introduce a simpler 
expression for the adiabatic evolution of 
Carter constant~\cite{newpaper}.  

\subsection{property of geodesics in Kerr}
We first discuss in a little more detail about geodesics in Kerr spacetime: 
$z^{\alpha}(\tau) =
(t_z(\tau),r_z(\tau),\theta_z(\tau),\varphi_z(\tau))$. 
Here $\tau$ is the proper time along the orbit.
Using a new parameter $\lambda$ defined by 
$d\lambda:=d\tau/\Sigma$, which was recently reintroduced 
by Mino in the context of radiation reaction~\cite{Mino03}, 
the geodesic equations are written as 
\begin{eqnarray}
&& \!\!\left(\frac{dr_z}{d\lambda}\right)^2 =
R(r_z), \qquad \left(\frac{d\cos\theta_z}{d\lambda}\right)^2 =
\Theta(\cos\theta_z),\label{eq:eom_r}\\
&& \!\!\!\!\frac{dt_z}{d\lambda} = 
-a(a{\cal E}\sin^2\theta_z-{\cal L})
+\frac{r_z^2+a^2}{\Delta}P(r_z),~
\frac{d\varphi_z}{d\lambda} =
-a{\cal E}+\frac{\cal L}{\sin^2\theta_z}
+\frac{a}{\Delta}P(r_z), \label{eq:eom_phi}
\end{eqnarray}
where
$
P(r)={\cal E}(r^2+a^2)-a{\cal L}$, $
R(r)=[P(r)]^2-\Delta[r^2+{\cal Q}]
$ and $
\Theta(\cos\theta)=
{\cal C} - ({\cal C}+a^2(1-{\cal E}^2)+{\cal L}^2)\cos^2\theta+
 a^2(1-{\cal E}^2)\cos^4\theta.
$
It should be noted that the equation for the $r$-component and
the one for the $\theta$-component are decoupled when we use 
$\lambda$.
The solutions of the first two equations (\ref{eq:eom_r}) are periodic. 
We denote the periods by $2\pi/\Omega_r$ and $2\pi/\Omega_\theta$, 
respectively. 
The other two equations (\ref{eq:eom_phi}) 
are integrated as 
\begin{eqnarray}
t_z(\lambda)&=&t^{(r)}(\lambda)+
   t^{(\theta)}(\lambda)+
   \left\langle {dt_z\over d\lambda}\right\rangle \lambda,\cr
\varphi_z(\lambda)&=&\varphi^{(r)}(\lambda)+
   \varphi^{(\theta)}(\lambda)+
   \left\langle {d\varphi_z\over d\lambda}\right\rangle \lambda, 
\end{eqnarray}
where $\langle \cdots \rangle$ 
means time average along the  
geodesic. 
$
t^{(r)}(\lambda) := \int d\lambda 
      \{(r_z^2+a^2)P(r_z) /\Delta
$
$
      -\langle(r_z^2+a^2)P(r_z) /\Delta\rangle\}
$
and 
$
t^{(\theta)}(\lambda) := -\int d\lambda 
      \{a(a{\cal E}\sin^2\theta_z-{\cal L})-
$$
            \langle a(a{\cal E}\sin^2\theta_z-{\cal L})\rangle\},  
$
are periodic functions with periods 
$2\pi/\Omega_r$ 
and $2\pi/ \Omega_\theta$, respectively. 
Functions $\varphi^{(r)}$ and
$\varphi^{(\theta)}$ are also defined in a similar way.

\subsection{simplified formula}
We start to simplify the expression of the formula for 
$d{\cal Q}/dt$. 
The self-force ${f}^{\alpha}$, which is 
defined by $u^\nu u^\mu_{~;\nu}=f^\mu$, is 
derived from the geodesic equation on a perturbed spacetime. 
Then $f^\mu$ is basically given by $-\delta\Gamma^\mu_{~\rho\sigma}
u^\rho u^\sigma$, where $\delta\Gamma^\mu_{~\rho\sigma}$ is 
the contribution to the Christoffel symbol from 
the metric perturbation ${\bf h}$. 
Taking into account the redefinition of the proper time 
so that $u^\mu u_\mu=-1$ is maintained, we obtain 
\[
f^{\mu}[{\bf h}]:=
-\frac{1}{2}(g^{\mu\nu}+u^{\mu}u^{\nu})
(h_{\nu\rho;\sigma}+h_{\nu\sigma;\rho}-h_{\rho\sigma;\nu})
u^{\rho}u^{\sigma}.
\]
Using this force, the evolution of Carter constant is given by 
\begin{eqnarray}
&&\frac{d{\cal Q}}{d\tau}
   =  2K_{\mu}^{\nu}u^{\mu}f_{\nu} \cr
&&\quad = 
 \lim_{x\rightarrow z}2\left[
K_{\mu}^{\nu}\tilde{u}^{\mu}
\partial_{\nu}{\Psi\over\Sigma}
-\frac{d}{d\tau}(K_{\mu}^{\nu}h_{\nu\alpha}
  \tilde{u}^{\alpha}\tilde{u}^{\mu})
+h_{\alpha\beta}\tilde{u}^{\alpha}\tilde{u}^{\mu}
(K_{\mu ;\nu}^{\beta}\tilde{u}^{\nu}
-K_{\mu}^{\nu}\tilde{u}^{\beta}_{;\nu})
\right].
\label{dqdt}
\end{eqnarray}
where 
$\Psi(x)=\Sigma \tilde u^\mu 
   \tilde u^\nu h_{\mu\nu}/2$ and 
$(\tilde u_t,\tilde u_r,\tilde
u_\theta,\tilde
u_\varphi):=(-{\cal E},\pm\sqrt{R(r)}/\Delta,$ 
$\pm\sqrt{\Theta(\cos\theta)}$ $/\sin\theta,{\cal L})$. 
This vector field $\tilde u_\mu$ 
is an extension of the four velocity of a particle 
in the sense that it satisfies 
$\tilde u_\mu(z(\lambda))=u_\mu(\lambda)$, but it 
differs from the parallel transport of the four velocity.  
Using the fact that 
$\tilde{u}_{r}$ and $\tilde {u}_\theta$, respectively, 
depend only on $r$ and $\theta$, we can easily verify the relation, 
$\tilde{u}_{\alpha;\beta}=\tilde{u}_{\beta;\alpha}$.

If we take the long-time average of Eq.~(\ref{dqdt}), 
the second term in the last line vanishes because it is a 
total derivative. 
Furthermore we can show 
that the long-time average of the third term also 
becomes higher order in $\mu$  
by using the relations 
$\tilde{u}_{\alpha;\beta}=\tilde{u}_{\beta;\alpha}$ and 
$K_{(\mu\nu;\rho)}$. 
Finally, we obtain 
\begin{equation}
\left\langle \frac{d{\cal Q}}{d\tau}\right\rangle =
\lim_{T\to\infty}{1\over T}\int_{-T}^T d\lambda \Sigma
 K_{\mu}^{\nu}\tilde{u}^{\mu}
\partial_{\nu}{\Psi(x)\over \Sigma}.
\label{dqdt2}
\end{equation}
We can derive an analogous expression more easily for 
the energy loss rate as~\cite{Gal'tsov82}
\begin{eqnarray}
\left\langle {d{\cal E}\over d\lambda}\right\rangle & = & 
\lim_{T\to\infty}{1\over 2T}\int_{-T}^T d\lambda \Sigma(-\eta_{(t)}^\alpha)
 f_\alpha [h_{\mu\nu}]\cr
 & = &  \lim_{T\to\infty}{1\over 2T}\int_{-T}^T 
   d\lambda \left[
       (-\eta_{(t)}^{\alpha})
       \partial_\alpha \Psi(x)\right]_{x\to z(\lambda)}, 
\label{shownbyGaltsov}
\end{eqnarray}
where $\eta_{(t)}^{\alpha}$ is the timelike Killing vector.

In any cases, all we need to know is $\Psi(x)$ for the 
radiative field. 
Using the formulas (\ref{hrad}) and (\ref{hrad3}), we find 
\begin{eqnarray}
\Psi^{(rad)}(x) = i
    \int {d\omega\over 2\pi\omega} \sum_{\ell,m} N^{in}
   \phi^{(in)}_{\omega,\ell,m}(x) 
   \int d\lambda' 
    \overline{\phi^{(in)}_{\omega,\ell,m}(z(\lambda'))}+\cdots,
\label{dEdlambda1}
\end{eqnarray}
where the normalization constant 
$N^{in}$ is defined by 
$N^{in}:=-2\pi\omega i 
    W({}_s R^{down}_\Lambda,{}_s R^{up}_\Lambda)$
$   {}_s \bar{\cal N}^{-1}\, 
 W({}_s R^{up}_\Lambda,{}_s R^{in}_{\Lambda})^{-1}
    W({}_s R^{down}_\Lambda,{}_s R^{out}_\Lambda)^{-1}
$. 
Hereafter we neglect the contribution 
from waves absorbed by black hole, since the extension is 
trivial.  

We will not repeat here the technical issues 
for further reduction of the formula discussed in Ref.~\citen{newpaper}. 
Instead, we just briefly mention essential points. 
First point is that the frequency $\omega$ is discretized as 
\begin{eqnarray}
\int d\lambda' 
    \overline{\phi^{(in)}_{\Lambda}(z(\lambda'))}
 & = & 
    \sum_{n_r,n_\theta} 
     2\pi \delta\left(\omega-\omega_m^{n_r,n_\theta}\right)
      {\bar Z_{\tilde\Lambda}},
\label{source}
\end{eqnarray}
with
\begin{equation}
\omega_m^{n_r,n_\theta}:=
\left\langle {dt_z/ d\lambda}\right\rangle^{-1}   
   \left(m\left\langle d\varphi_z/d\lambda\right\rangle
        +n_r \Omega_r + n_\theta\Omega_\theta\right).
\end{equation}
Here $\tilde\Lambda$ represents a set of eigenvalues 
$\{\ell,m,n_r,n_\theta\}$. 
For $\left\langle{d{\cal E}/dt}\right\rangle$, we obtain   
\begin{equation}
 \left\langle{d{\cal E}\over dt}\right\rangle 
    = -\sum_{\tilde\Lambda}
       \vert Z_{\tilde\Lambda}\vert^2, 
\label{standard}
\end{equation} 
and similarly, 
\begin{equation}
\left \langle{d{\cal L}\over dt}\right\rangle 
    =  -\sum_{\tilde\Lambda}
       {m\over \omega_m^{n_r,n_\theta}}
       \vert Z_{\tilde\Lambda}\vert^2. 
\label{standardLz}
\end{equation}
In the above, we have set $2N^{in}=1$ by rescaling the 
amplitude of radial functions ${}_s R_{\Lambda}$ appropriately. 
Although it is not manifest from our definition of $N^{in}$, 
we can show that it is real.  
For $\langle d{\cal Q}/dt\rangle$, 
writing down the expression in Eq.~(\ref{dqdt2}) explicitly, 
we arrive at 
\begin{eqnarray}
\int & d\lambda & \left[\Sigma 
    K^\nu_\mu\tilde u^\mu\partial_\nu{\Psi^{(rad)}(x)\over\Sigma}
       \right]_{x=z(\lambda)}\cr
  & = & \int d\lambda \left[
       \left(-{P(r)\over \Delta}
            ((r^2+a^2)\partial_t+a\partial_{\varphi})
            - {dr_z\over d\lambda}\partial_r\right)
            \Psi^{(rad)}(x)\right]_{x=z(\lambda)}.
\label{CarterReaction}
\end{eqnarray}
Using Eqs.(\ref{dEdlambda1}) and (\ref{source}), 
the above expression reduces to 
\begin{eqnarray}
\Re\left\{
      \sum_{\tilde \Lambda} i        
        \bar Z_{\tilde \Lambda}
     \int d\lambda \left[
       \left(-{P(r)\over \Delta}
            ((r^2+a^2)\partial_t+a\partial_{\varphi})
            - {dr_z\over d\lambda}\partial_r\right)
            \phi_{\tilde\Lambda}(x)\right]_{x=z(\lambda)}\right\} .
\label{CarterReaction}
\end{eqnarray}
The second point is to notice that the integrand is now
a double periodic function with 
periods $2\pi/\Omega_r$ and $2\pi/\Omega_\theta$.
Suppose $f$ is a function of $g^{(r)}$ and $g^{(\theta)}$, and 
$g^{(r)}(\lambda)$ and $g^{(\theta)}(\lambda)$ are periodic 
with periods $2\pi/\Omega_r$ and $2\pi/\Omega_\theta$, respectively. 
Then, in general, 
\begin{eqnarray}
 &&\lim_{T\to\infty}{1\over 2T}\int_{-T}^{T} 
    d\lambda f(g^{(r)}(\lambda),g^{(\theta)}(\lambda))\cr
  &&\quad 
 =   {\Omega_r\Omega_\theta\over (2\pi)^2}\int_{0}^{2\pi\Omega_r^{-1}} 
        \hspace{-3mm}d\lambda_r
    \int_{0}^{2\pi\Omega_\theta^{-1}}
       \hspace{-3mm} d\lambda_\theta 
           f(g^{(r)}(\lambda_r),g^{(\theta)}(\lambda_\theta)), 
\end{eqnarray}
holds. 
Using this formula, the integral (\ref{CarterReaction}) 
can be integrated by parts. 
Then finally we arrive at the formula 
\begin{eqnarray}
\left\langle {d{\cal Q}\over dt}\right\rangle
  =  2\left\langle {(r^2+a^2)P\over \Delta}\right\rangle
         \left\langle{d{\cal E}\over dt}\right\rangle
        -2\left\langle {a P\over \Delta}\right\rangle
           \left\langle{d{\cal L}\over dt}\right\rangle
            + 2
           \sum_{\tilde\Lambda}
           {n_r\Omega_r\over \omega_m^{n_r,n_\theta}} 
                \vert Z_{\tilde\Lambda}\vert^2. 
\label{dQdlambda}
\end{eqnarray}
This expression is as easy to evaluate as $\langle d{\cal E}/dt \rangle$ 
and $\langle d{\cal L}/dt \rangle$. 
To evaluate the last term, we have 
only to replace $m$ 
with $n_r \Omega_r$ in 
the expression for $\langle d{\cal L}/dt \rangle$, (\ref{standardLz}). 

Although the final expressions 
for $\langle d{\cal L}/dt\rangle$ and 
$\langle d{\cal Q}/dt\rangle$ are quite similar, 
there is a big difference between them. 
The squared amplitude of each partial wave, 
$|Z_{\tilde\Lambda}|^2$, is a measurable quantity in the 
asymptotic regions, i.e., near the future null infinity or the future 
event horizon. The eigenvalues $\omega$ and $m$ 
also can be read from the waveform in the asymptotic region. 
Hence, the expressions for $\langle d{\cal E}/dt\rangle$ 
and $\langle d{\cal L}/dt\rangle$ 
are solely written in terms of the asymptotic waveform. 
This is consistent with the fact that the balance argument 
applies for these quantities. 
On the other hand, the eigenvalue $n_r$, 
which appears in the expression for 
$\langle d{\cal Q}/dt\rangle$, 
is not a quantity which can be read from the asymptotic waveform 
without knowing the particle orbit. We can read 
the frequencies from the asymptotic waveform, but the frequency 
itself does not tell the numbers $n_r$ and $n_\theta$, 
without an input of additional information about the orbit. 
This is consistent with our understanding that the balance argument 
cannot be used for evaluating $\langle d{\cal Q}/dt\rangle$.

\subsection{long period orbital evolution}
\label{longperiod}
One may ask how we can use the knowledge about the adiabatic 
evolution of ${\cal Q}$ to evaluate the long time orbital evolution. 
Just to be consistent with a given set of constants of motion $I^i$, 
we can choose $r, \theta$ and $\varphi$ arbitrarily at each time $t$ 
as far as $r$ and $\theta$ are within the allowed range. 
Hence, the evolution of the constants of motion is not complete 
at all as a description of the orbital evolution. 
Nevertheless, 
combined with the normalization of four velocity, 
\begin{equation}
 g^{BH}_{\mu\nu} 
   {~~~dz^\mu\over \Sigma d\lambda} {~~~dz^\nu\over \Sigma d\lambda}=-1, 
\end{equation}
three constants of motion are sufficient to specify the four velocity 
as a function of $r$ and $\theta$. 
Even if we take into account the instantaneous self-force, 
the equations of motion integrated once 
((\ref{eq:eom_r}) and (\ref{eq:eom_phi})) are kept unchanged 
since they are merely algebraic relations between $u^\mu$ and 
$I^i$. 
Hence, 
if the evolution of the ``constants of motion'' $I^i$ is given,  
we do not need further information about the self-force 
in order to evolve $(r(\lambda), \theta(\lambda), 
t(\lambda), \varphi(\lambda)$). 
Here we give a prescription how to 
integrate the orbit for a long period of time. 
The properties of orbits in 
Kerr spacetime which we have already mentioned 
largely simplify the problem of solving the orbital evolution. 
We will find that the leading order approximation can be obtained by 
using the adiabatic approximation as expected. 
We will also find that the leading 
order corrections to the adiabatic approximation 
coming from the instantaneous self-force in the linear order 
are given by 
a few time-averaged quantities constructed from the self-force. 
The basic idea of the discussion in this subsection
is found in Ref.~\citen{Mino05}.

We begin with discussing orbits for fixed constants of motion $I^i$. 
It is convenient to introduce phase functions  
$\chi^r$ and $\chi^\theta$ by $\chi^a=\Omega_a(I^i)\lambda$, 
where $\Omega_r(I^i)$ and $\Omega_\theta(I^i)$ are 
angular frequencies of oscillations in $r$ and $\theta$ directions 
respectively. 
Note that we can choose the initial values of the phases 
$\chi^a(0)$ 
arbitrarily. 
We define functions 
$\hat r(I^i,\chi^r)$ and $\hat \theta(I^i,\chi^\theta)$
as solutions of the $r$-and $\theta$-components of the 
geodesic equations. 
Here we fix the ambiguity in the choice of phases of these functions 
so as to satisfy 
$\chi^r=0$ mod $2\pi$ for $\hat r(\chi^r)=r_{-}$ and 
$\chi^\theta=0$ mod $2\pi$ for $\hat \theta(\chi^\theta)=\theta_{-}$.
Here $r_-$ and $\theta_-$ are minima of $r$ and $\theta$ for 
given $I^i$.  
Since the evolutions of $r$ and $\theta$ for $0<\chi^a <\pi$ 
and for $\pi<\chi^a <2\pi$ are symmetric, we automatically have 
$\chi^r=\pi$ mod $2\pi$ for $\hat r(\chi^r)=r_{+}$ and 
$\chi^\theta=\pi$ mod $2\pi$ for $\hat \theta(\chi^r)=\theta_{+}$  
with this choice of phases.  
Further we redefine the functions 
$t^{(r)}$ and $\varphi^{(r)}$ here as 
functions of $I^i$ and $\chi^r$, and similarly 
$t^{(\theta)}$ and $\varphi^{(\theta)}$ as  
functions of $I^i$ and $\chi^\theta$. 
Again, for definiteness we fix these functions so that 
$t^{(a)}(I^i,\chi^a)=0$ and $\varphi^{(a)}(I^i,\chi^a)=0$ 
for $\chi^a=0$ mod $2\pi$. 

Now we are ready to introduce our parametrization to 
describe orbits when we take into account the self-force. 
Our proposal is to promote $I^i$ and $\chi^a$ 
to functions of $\lambda$ as 
\begin{eqnarray}
 r(\lambda) & = & \hat r(I^i(\lambda),\chi^r(\lambda)), \cr
 \theta(\lambda) 
   & = & \hat \theta(I^i(\lambda),\chi^\theta(\lambda)), \cr
 t(\lambda) & = & t^{(r)}(I^i(\lambda),\chi^r(\lambda))
        +t^{(\theta)}(I^i(\lambda),\chi^\theta(\lambda))
        +\tilde\chi^t(\lambda),\cr
 \varphi(\lambda) & = & \varphi^{(r)}(I^i(\lambda),\chi^r(\lambda))
        +\varphi^{(\theta)}(I^i(\lambda),\chi^\theta(\lambda))
        +\tilde\chi^\varphi(\lambda).
\label{adiabaticsol}
\end{eqnarray}
Here we have also introduced $\tilde\chi^A(\lambda)$ instead of the 
integrals such as 
$\int \left\langle {dt/ d\lambda}
\right\rangle (I^i(\lambda))d\lambda$. 
Below we derive equations for 
$I^i(\lambda)$, $\chi^a (\lambda)$ and $\tilde\chi^A(\lambda)$. 

First we examine the evolution equation for $r$.
As we have mentioned earlier, 
Eqs.~(\ref{eq:eom_r}) and (\ref{eq:eom_phi})) 
are kept unchanged even if the self-force is taken into account. 
Therefore we have
\begin{equation}
 {dr \over d\lambda}=
   {\partial \hat r(I^i(\lambda),\chi^r)\over \partial \chi^r}
    \Omega_r(I^i(\lambda)), 
\end{equation}
since $\hat r$ is the solution of 
the geodesic equation for fixed $I^i$. On the other hand, 
taking the $\lambda$-derivative of $r(\lambda)$ in the form 
given in Eq.~(\ref{adiabaticsol}), we obtain 
\begin{equation}
 {d\over d\lambda}
  \hat r(I^i(\lambda),\chi^r(\lambda))=
   {\partial \hat r\over \partial I^i}{dI^i\over d\lambda}+
   {\partial \hat r\over \partial \chi^r}{d\chi^r\over d\lambda}. 
\end{equation}
Comparing these two expressions, 
we obtain an equation for $d\chi^r/d\lambda$ as 
\begin{equation}
 {d\chi^r\over d\lambda}
  =\Omega_r+\delta\left({d\chi^r\over d\lambda}\right), 
\label{eqchir}
\end{equation}
where 
\begin{equation}
 \delta\left({d\chi^r\over d\lambda}\right)
  = -\left({\partial \hat r\over \partial\chi^r}\right)^{-1}
       {\partial \hat r\over \partial I^i}{dI^i\over d\lambda}. 
\end{equation}
Near the turning points $r=r_{\pm}$, $\hat r$ can be expanded as 
$\hat r=r_\pm(I^i)+O((\Delta\chi^r)^2)$, where $\Delta\chi^r$ is 
the difference of $\chi$ from its value at $r=r_{\pm}$. 
Then $\delta\left({d\chi^r/d\lambda}\right)$ 
looks singular since 
${\partial \hat r/\partial\chi^r}$ behaves 
like $\approx \Delta\chi$ near the turning points, but 
in fact this term is not singular. 
As we shall see immediately below, 
the other factor 
$({\partial \hat r/\partial I^i})({dI^i/d\lambda})$
simultaneously goes to $0$ at the turning points 
as long as the self-force stays finite. 

We shall show $({\partial \hat r/\partial I^i})({dI^i/d\lambda})
=O((\Delta \chi^r)^2)$. 
Differentiating the equation $(dr/d\lambda)^2=R(I^i,r)$ 
with respect to $\lambda$, we have 
\begin{equation}
 2\left({dr\over d\lambda}\right){d^2r\over d\lambda^2}
 = {\partial R\over\partial I^i}{dI^i\over d\lambda}
   +{\partial R\over\partial r}{dr \over d\lambda}.
\end{equation}
Hence, we conclude that $({\partial R(I^i,r)/\partial I^i})
    ({dI^i/d\lambda})=0$ for $r=r_{\pm}$. 
On the other hand, by the definition of $r_{\pm}$, 
we have $R(I^i,r_{\pm}(I^i))\equiv 0$. 
Differentiating this identity with respect to $I^i$, and 
contracting it with $dI^i/d\lambda$, 
we have
\begin{equation}
  \left({\partial R\over \partial r} \right)_{r=r_\pm}
  {\partial r_\pm \over \partial I^i}{dI^i\over d\lambda} 
   =
   -\left({\partial R\over\partial I^i} \right)_{r=r_\pm}
    {dI^i\over d\lambda} =0.
\end{equation}
Hence, except for circular orbits, in which 
$\left({\partial R/\partial r} \right)_{r=r_\pm}=0$, 
we establish 
$({\partial r_\pm/ \partial I^i})$ $({dI^i/ d\lambda})=0$.  
When the orbit is circular, $r$ is given as a function of $I^i$. 
Hence, we do not have to care about the evolution of 
$\chi^r$. 

Now we return to Eq.~(\ref{eqchir}). 
To evaluate $dI^i/d\lambda$, here we accept the use of 
the geodesic momentarily tangential to the orbit in  
evaluating the self-force. The errors caused by this approximation need 
future investigation. 
Under this approximation, 
the change of the ``constants of motion'' 
becomes a double periodic function in $\lambda$, and is expanded as 
\begin{equation}
{dI^i\over d\lambda}=\left\langle{dI^i\over d\lambda}\right\rangle
   +\!\!\!\!\sum_{(n_r,n_\theta)\ne(0,0)} \!\!\!\!
     J_{n_r,n_\theta}^i(I^j(\lambda)) 
     e^{i(n_r\chi^r(\lambda)+n_\theta\chi^\theta(\lambda))}+O(\mu^2),   
\label{evolI}
\end{equation}
where $J_{n_r,n_\theta}(I^j(\lambda))$ are coefficients 
to be computed from the instantaneous self-force. 
The last term of $O(\mu^2)$ is the correction due to the second order
self-force, which currently we do not know how to evaluate. 
In what follows we will give a rough estimate of the error due to this 
second order contribution, but its derivation is not rigorous 
at all. 
The second term in (\ref{eqchir}) is also a regular 
double periodic function. Hence it accept a similar expansion. 
We can write Eq.~(\ref{eqchir}) as 
\begin{eqnarray}
 {d\chi^r\over d\lambda}
  &=& \Omega_r
   + \left\langle \delta\left({d\chi^r\over d\lambda}\right)
     \right\rangle (I^i(\lambda)) \cr 
  &&  +\sum_{(n_r,n_\theta)\ne(0,0)} 
     \delta\Omega_r^{n_r,n_\theta}(I^j(\lambda)) 
     e^{i(n_r\chi^r(\lambda)+n_\theta\chi^\theta(\lambda))}+O(\mu^2).  
\label{evolvechi}
\end{eqnarray}
The effect of the third term
looks as large as the second one at first sight. However, 
if we integrate the above equation once, one finds that 
the second time grows as $O(\mu\Delta \lambda)$, 
where $\Delta\lambda$ is the length of integration time. 
Since we are interested in long term orbital evolution, 
$\Delta\lambda$ is assumed to be large. 
On the other hand, the third term can be integrated by parts as 
\begin{eqnarray}
\int d\lambda &&\delta\Omega_r^{n_r,n_\theta}
           (I^j(\lambda)) 
     e^{i(n_r\chi^r(\lambda)+n_\theta\chi^\theta(\lambda))}
     \cr
&& =
      {\delta\Omega_r^{n_r,n_\theta}(I^j(\lambda)) 
       \over i (n_r(d\chi^r/d\lambda)+
                n_\theta(d\chi^\theta/d\lambda))}
     e^{i(n_r\chi^r(\lambda)+n_\theta\chi^\theta(\lambda))}
\cr
&& \quad -  \int d\lambda 
     e^{i(n_r\chi^r(\lambda)+n_\theta\chi^\theta(\lambda))} 
         {dI^j\over d\lambda}{d\over dI^j}
         {\delta\Omega_r^{n_r,n_\theta}(I^j(\lambda)) 
          \over i (n_r(d\chi^r/d\lambda)+
                n_\theta )(d\chi^\theta/d\lambda)}. 
\label{deltaOmega}
\end{eqnarray}
The first term in the last expression does not grow even 
for a large value of $\Delta\lambda$, 
while the second term becomes higher order in $\mu$ due to 
the appearance of the factor $dI^i/d\lambda$. 
We express the result of integration of (\ref{evolvechi}) as 
\begin{equation}
\chi^r(\lambda)=
\int \left\{
  \Omega_r+\left\langle \delta\left({d\chi^r\over
         d\lambda}\right) \right\rangle\right\}  d\lambda 
    +O(\mu(\Delta\lambda)^0) 
      +O(\mu^2\Delta\lambda). 
\end{equation}
The last term is the correction coming from the second order 
self-force. 
Discussion about the $\theta$-component is 
completely parallel to the $r$-component. 

Next we consider the $t$-component. By the definitions of 
$t^{(r)}$ and $t^{(\theta)}$, we have 
\begin{equation}
 {dt\over d\lambda}=
  {\partial t^{(r)}\over \partial \chi^r}
  \Omega_r
  +  {\partial t^{(\theta)}\over \partial \chi^\theta}
  \Omega_\theta+\left\langle{dt\over d\lambda}\right\rangle. 
\end{equation}
On the other hand, 
differentiating $t(\lambda)$ in the form 
of (\ref{adiabaticsol}), we obtain 
\begin{equation}
 {dt\over d\lambda}=
 {\partial t^{(r)}\over \partial I^i}{dI^i\over d\lambda} 
  +{\partial t^{(r)}\over \partial \chi^r}{d\chi^r\over d\lambda} 
  + {\partial t^{(\theta)}\over \partial I^i}{dI^i\over d\lambda} 
  +{\partial t^{(\theta)}\over \partial \chi^\theta}{d\chi^\theta\over d\lambda} 
   +{d\tilde\chi^t\over d\lambda}.
\end{equation}
From a comparison of these two equations, we find 
that the evolution of $\tilde\chi^t$ is to be determined by 
\begin{eqnarray}
 {d\tilde\chi^t\over d\lambda}
 &=& \left\langle{dt\over d\lambda}\right\rangle(I^j(\lambda)) 
  +\left\langle\delta\left({d\tilde\chi^t\over d\lambda}\right)
        \right\rangle(I^j(\lambda)) \nonumber \\
&&   ~~~~+\!\!\!\!\sum_{(n_r,n_\theta)\ne(0,0)} 
     \left({d\tilde\chi^t\over d\lambda}\right)^{n_r,n_\theta}(I^j(\lambda)) 
     e^{i(n_r\chi^r(\lambda)+n_\theta\chi^\theta(\lambda))}+O(\mu^2), 
\label{evolvexi}
\end{eqnarray}
with 
\begin{equation}
\delta\left({d\tilde\chi^t\over d\lambda}\right):=
 -{\partial t^{(r)}\over \partial I^i}{dI^i\over d\lambda} 
  -{\partial t^{(r)}\over \partial \chi^r}\delta
   \left({d\chi^r\over d\lambda} \right)
 -{\partial t^{(\theta)}\over \partial I^i}{dI^i\over d\lambda} 
  -{\partial t^{(\theta)}\over \partial \chi^\theta}\delta
   \left({d\chi^\theta\over d\lambda} \right). 
\end{equation}
Then, this equation can be integrated as in the case of $\chi^a$. 
A completely parallel discussion goes through 
for the $\varphi$-component, too. 

Now we consider the evolution of $I^i(\lambda)$. 
The evolution equations for $I^i(\lambda)$ are already given by 
Eq.~(\ref{evolI}). 
We express the result of integration of (\ref{evolI}) as 
\begin{equation}
I^i(\lambda)=I^i_{Ad}(\lambda)+\delta I^i(\lambda).
\end{equation}
The first term $I^i_{Ad}(\lambda)$ is the contribution 
only from the first term on the right hand side of 
Eq.~(\ref{evolI}). We find that the contribution to 
$\delta I^i(\lambda)$ from the terms including $J_{n_r,n_\theta}^i$
remains $O(\mu(\Delta\lambda)^0)$ or higher.  
On the other hand, the second order self-force contribute
to $\delta I^i$ as terms of $O(\mu^2\Delta\lambda)$. 

The errors in $I^i(\lambda)$ propagate to 
$\chi^a(\lambda)$ and $\tilde\chi^A(\lambda)$. 
By using the same argument that we 
have already used many times, the propagated errors caused by 
the terms which include $J_{n_r,n_\theta}^i$ are at most 
$O(\mu(\Delta\lambda)^0)$, 
while the effect of second order self-force can be as large as 
$O(\mu^2\Delta\lambda^2)$ for both 
$\chi^a(\lambda)$ and $\tilde\chi^A(\lambda)$. 
To conclude, we found 
\begin{eqnarray}
I^i(\lambda) & = & I^i_{Ad}(\lambda)+O(\mu(\Delta\lambda)^0)
         +O(\mu^2\Delta\lambda),\cr
\chi^a(\lambda) & = & 
  \chi^a_{Ad}(\lambda)+\int 
   \left\langle \delta\left({d\chi^r\over d\lambda}\right)
     \right\rangle d\lambda
    +O(\mu(\Delta\lambda)^0)
         +O(\mu^2\Delta\lambda^2),j\cr
\tilde\chi^A(\lambda) & = & 
  \tilde\chi^A_{Ad}(\lambda)+
   \int\left\langle\delta\left({d\tilde\chi^A\over d\lambda}\right)
      \right\rangle d\lambda
     + O(\mu(\Delta\lambda)^0)
         +O(\mu^2\Delta\lambda^2), 
\end{eqnarray}
with
\begin{eqnarray}
 && I^i_{Ad}(\lambda):=\int\left\langle
      {dI^i\over d\lambda}\right\rangle(I^j_{Ad}(\lambda))\, d\lambda,\cr
&& \chi^a_{Ad}(\lambda):=\int \Omega_a(I_{Ad}^i(\lambda)) d\lambda,\cr
&&  \tilde\chi^t_{Ad}(\lambda):=\int \left\langle
      {dt\over d\lambda}\right\rangle (I_{Ad}^i(\lambda)) d\lambda, 
\quad
 \tilde\chi^\varphi_{Ad}(\lambda):=\int \left\langle
      {d\varphi\over d\lambda}\right\rangle (I_{Ad}^i(\lambda)) d\lambda. 
\end{eqnarray}
The last term for each $I^i(\lambda)$, 
$\chi^a(\lambda)$ and 
 $\tilde\chi^A(\lambda)$ comes from the second order self-force. 
The second terms in the expressions for $\chi^a(\lambda)$ and 
 $\tilde\chi^A(\lambda)$ are $O(\mu\Delta\lambda)$. 
When $\mu\Delta\lambda\ll 1$, the deviation from 
the adiabatic approximation is small. 
In this sense adiabatic approximation is a good approximation. 
On the other hand, once $\mu\Delta\lambda$ becomes $O(1)$, 
we cannot neglect the second order self-force, either. 
Hence, roughly speaking, the orbital evolution which takes into 
account the instantaneous self-force but only at the linear order 
will not be a better approximation than that 
obtained by using the adiabatic approximation. Even when 
$\mu\Delta\lambda$ is moderately small, the leading corrections 
are given by quantities averaged over a long period of time, 
$\left\langle \delta\left({d\chi^r/ d\lambda}\right)
     \right\rangle$ and 
$\left\langle\delta\left({d\tilde\chi^A/d\lambda}\right)
      \right\rangle$. Again we can repeat the 
same argument that was used to justify 
replacing the retarded field with the radiative one 
when we evaluate $d{\cal Q}/dt$. 
Therefore adiabatic approximation 
is sufficient to calculate these averaged quantities. 

Nevertheless, we are not saying that the study of the 
instantaneous self-force 
is not important. 
When we consider the whole process of inspiral of binaries, 
the time scale is inversely proportional to $\mu$. Therefore, 
both the terms of $O(\mu\Delta\lambda)$ and 
of $O(\mu^2\Delta\lambda^2)$ 
potentially cause measurable effects. 
To evaluate these effects, we need to know 
(maybe some time average of) the second order self-force, 
and for this purpose the study of the instantaneous 
self-force at linear order 
will be necessary. Furthermore, 
there might be situations in which the 
effects of the oscillating part of the self-force, 
i.e., the terms which contain $J^i_{n_r,n_\theta}$ 
are significantly enhanced. In the above discussion, 
when we perform integration by parts such 
as shown explicitly in (\ref{deltaOmega}), a factor 
$n_r(d\chi^r/d\lambda)+
                n_\theta(d\chi^\theta/d\lambda)$ 
appeared in the denominator. As the ``constants of motion'' $I^i$ 
evolve, this denominator eventually 
can cross $0$ for a certain combination 
of $n_r$ and $n_\theta$.
In such a case, the contribution 
from $J^i_{n_r,n_\theta}$ may leave some significant effect 
on the orbital evolution\cite{Mino05}. 
Suppose that 
$n_r (d\chi^r/d\lambda)
+n_\theta(d\chi^\theta/d\lambda)$ vanishes 
at $\lambda=\lambda_0$. 
Such a stationary phase point (SPP)
will cause an additional shift in $I^i(\lambda)$ which 
will be estimated by the Gaussian integral 
\begin{eqnarray}
 \Delta I^i_{SPP} 
    &\approx &\int d\lambda
     J_{n_r,n_\theta}^i(I^j(\lambda)) 
     e^{i(n_r\chi^r(\lambda_0)+n_\theta\chi^\theta(\lambda_0))
       +{i\over 2}\{n_r (d^2\chi^r/d\lambda^2)
           +n_\theta(d^2\chi^\theta/d\lambda^2)\}(\lambda-\lambda_0)^2}
\cr
    &\approx &
     \sqrt{2\pi\over i
       (n_r {d\Omega_r\over dI^j}{dI^j\over d\lambda}
           +n_\theta{d\Omega_\theta\over dI^j}{dI^j\over 
                      d\lambda})}
    J_{n_r,n_\theta}^i(I^j(\lambda_0)) 
     e^{i(n_r\chi^r(\lambda_0)+n_\theta\chi^\theta(\lambda_0))}. 
\end{eqnarray}
This correction is $O(\sqrt{\mu})$, and it will induce shifts 
in phases $\chi^a$ and $\tilde\chi^A$ of $O(\sqrt{\mu}\Delta\lambda)$. 
Hence, this naive order counting indicates that this correction is 
the leading order correction to the adiabatic approxiamtion. 
However, when we consider relatively 
non-relativistic orbits with small eccentricity, such stationary 
points will appear only when the values of $n_r$ 
and/or $n_\theta$ are large. In such cases  
the coefficients $J_{n_r,n_\theta}^i$ will be suppressed 
by some large powers of the eccetricity or $(v/c)$. 
Hence, the effect will practically remain small. 
To the contrary, when we consider highly eccentric 
or highly relativistic orbits, the corrections 
due to stationary points may become really
$O(\sqrt{\mu}\Delta\lambda)$ without any significant 
additional suppression. Again, to evaluate these effect 
quantitatively, we need to compute the instantaneous self-force. 

Before closing this subsection, 
we want to emphasize that the errors in phases 
$\chi^a$ and $\tilde\chi^A$ can grow in a dynamical time scale 
if we do not use an appropriate 
integration scheme as presented here. 
Namely, they can be as large as $O(\mu
(e^{\alpha\Delta\lambda}-1))$ with $\alpha$ being 
a constant of $O(1)$.

\section{Toward post-Teukolsky formalism}
Now we consider to extend the Teukolsky formalism 
to the second order, which seems to be indispensable to 
improve predictions for waveforms beyond the level 
of adiabatic approximation. 
Perturbed Einstein equations up to the second order 
take the form 
\begin{equation}
\delta G_{\mu\nu}\left[{\bf h}^{(2)}\right]
  = T^{(2)}_{\mu\nu}
   -G^{[2]}_{\mu\nu}\left[{\bf h}^{(1)},{\bf h}^{(1)}\right],  
\end{equation}
where $\delta G_{\mu\nu}$ is nothing but 
$G^{[1]}_{\mu\nu}$ in Eq.~(\ref{Gexpand}). 
Projection of this equation can be done formally 
as in the case of linear perturbation as 
\begin{equation}
 L\zeta^{(2)}=\sqrt{-g}T^{(2)}.
\end{equation}
Here $\zeta^{(2)}$ is defined in the same way as $\zeta^{(1)}$ 
just substituting ${\bf h}^{(1)}$ with ${\bf h}^{(2)}$. 
The second order source term $T^{(2)}$ has a spatially 
extended distribution due to the non-linearity of gravity. 
The differential operator $L$ is the same as the one for the 
linear perturbation. Therefore all the technical difficulties 
which arise in the second order for the first time 
are in how to evaluate the source term. 
To obtain the corrections to 
the energy momentum tensor $T^{(2)}_{\mu\nu}$, 
we need to know the correction 
to the trajectory of the particle taking into account the 
self-force. Hence as a first step toward the post-Teukolsky 
formalism, we examine the self-force. 

\subsection{gravitational self-force}
When we consider the point particle limit, the full
self-force diverges at the location of the particle, and hence
needs to be regularized. It is known that the properly regularized
self-force is given by the tail part of the self-field, 
which is obtained by subtracting the direct part 
from the full field. 
Here we do not give the precise definition of the direct part, 
but, roughly speaking, it is the part that propagates along the 
light cone unaffected by curvature scattering. 
Since direct part does not contain curvature scattering effect, 
the direct part of the local field near the particle is 
solely determined by the local geometry and 
the orbital elements of the particle.  
The justification of this
prescription is given in \cite{DB} for the scalar and
electro-magnetic cases, and in \cite{MST97,QW97} for the gravitational
case. An equivalent but more elegant decomposition of the Green
function was proposed by Detweiler and Whiting \cite{DW03,detweiler}, 
in which the direct part is replaced by the $S$-part and the tail
part by the $R$-part. The $S$-part is defined so as to 
vanish when two arguments $x$ and $x'$ are timelike.  
When $S$-part is subtracted from the
full field, the remainder $R$-part gives the regularized self-force. 
The advantage of this new decomposition is that
the $S$-part is symmetric with respect to $x$ and $x'$, and it
satisfies the same equation as the retarded Green function does.
This implies that the $R$-part now satisfies the source-free,
homogeneous equation.

\subsection{subtraction and regularization\cite{nakano}
}

Since we do not know a direct way to compute the 
$R$-part, we instead compute
$
 F^\alpha\left[\psi^R \right](\tau)
  =F^\alpha\left[\psi^{full} \right](\tau)
    - F^\alpha\left[\psi^S \right](\tau).$
Since both the terms on the right hand side 
are divergent, this expression does not make sense 
unless we regularize the divergent quantities. 
A practical way of regularization 
is the so-called ``point splitting'' regularization. 
We evaluate the expression for the 
force not exactly at the location of the particle but 
slightly off the point. If we subtract the 
$S$-part appropriately, the coincidence limit must be 
well-defined. 

{}For an actual computation, 
we decompose divergent {\it full}- and $S$-parts of the force into 
terms labelled by the total angular momentum 
$\ell$~\cite{Barack:1999wf,Burko:2000xx,MNS,NMS,BMNOS,Barack:2002mh,Detweiler:2002gi,Barack:2002bt,Barack:2003mh}, 
\begin{equation}
 F^\alpha\left[\psi^R \right](\tau)
  =\lim_{x\to z(\tau)}\sum_{\ell=0}^{\infty}
      \left(
          F_{\ell}^\alpha\left[\psi^{full} \right](\tau)
         - F_{\ell}^\alpha\left[\psi^S \right](\tau)\right). 
\end{equation}
Then each $\ell$-order term stays finite even in the 
coincidence limit. 
Here the question is whether we can change the order of 
the two operations as 
\begin{equation}
 F^\alpha\left[\psi^R \right](\tau)
  =\sum_{\ell=0}^{\infty} \lim_{x\to z(\tau)}
      \left(
          F_{\ell}^\alpha\left[\psi^{full} \right](\tau)
         - F_{\ell}^\alpha\left[\psi^S \right](\tau)\right). 
\end{equation}
In general these two operations 
do not commute. 

For example, let us 
define two functions $A(x)$ and $B(x)$ 
as $A(x)=\sum_{\ell=0}^\infty (1-x)^\ell $
and $B(x)=\sum_{\ell=1}^\infty (1-x)^\ell $ 
for $x\geq 0$. 
For $x\ne 0$, both $A(x)$ and $B(x)$ are convergent. 
Therefore we can compute the expression $A(x)-B(x)$ unambiguously, 
and it is $1$.  
In this case it is trivial that 
the difference $A(x)-B(x)=1+\sum_{\ell=1}^\infty 0$ 
is uniformly convergent, and each $\ell$-th order term is continuous. 
Therefore, one can change the order of two operations. 
Irrespectively of the order of operations we arrive at 
the same answer. 
The importance of the conditions of uniform convergence 
will become clearer if we consider the case with 
$B(x)=\sum_{\ell=1}^\infty (1-x)^{\ell-1}$. In this case $A(x)-B(x)=0$. 
Nevertheless, if we take the limit $x\to 0$ first, we end up with 
$A(0)-B(0)=1$. Now the term at the $\ell$-th order is 
$(1-x)^\ell-(1-x)^{\ell-1}=-x(1-x)^{\ell-1}$. As $x$ becomes closer 
to $0$, the convergence becomes worse. This series is convergent 
at each point $x$ but not uniformly. 

A practical way to guarantee the uniform convergence 
in the harmonic expansion of the force
is to use the same prescription for both 
{\it full}-part and $S$-part to extend 
the force off the trajectory and to use the same 
harmonics.   
The $S$-part of the force is determined by the local expansion near 
the particle. It is composed of terms like 
\begin{equation}
\sum { R^b \Theta^c \Phi^d\over \epsilon^a }
 f_{abcd}[z^\mu(\tau),u^\mu(\tau)],
\end{equation}
Here, the separation from the trajectory of the particle
$(x-z(\tau))^\mu$ was denoted by $(T,R,\Theta,\Phi)$, 
and $\epsilon$ is the spatial distance between $x^\mu$ and the 
trajectory. This type of function can be expanded 
in terms of spherical harmonics~\cite{Barack:1999wf,MNS,BMNOS}. 
On the other hand, the ${\it full}$-part composed of the 
master variable is also computed 
by using the spherical harmonic decomposition\footnote{
The way how to extend the force must be carefully chosen 
so that the harmonic decomposition of the {\it full}-part of the 
force is computable\cite{Barack:1999wf}. Alternative way to avoid 
this problem is to subtract $S$-part 
at the level of the metric perturbations\cite{NSS03} 
(or at the level of the master variable\cite{Sanjay}).}. 
After averaging over the angular direction from 
the location of the particle, 
we will obtain the {\it full}- and the 
$S$-parts of the force in the form of
\begin{equation} 
F^\alpha(\tau,R,X)=
\sum_{\ell=0}^{\infty}
F_{\ell}^\alpha(\tau,R)P_\ell(\cos X), 
\end{equation}
where $X$ is the angle between $x^\mu$ and $z^\mu(\tau)$. 
These expansions have 
a little ambiguity since the extension of 
the self-force is given only locally, in the vicinity of 
the particle. 
Nevertheless, 
the ambiguity does not affect 
the force evaluated near the particle 
after summation over $\ell$ since 
that is the basic requirement for the construction of 
harmonic expansion~\cite{Barack:1999wf,MNS,BMNOS}. 

Now we consider the difference between the {\it full}-part and 
the $S$-part of the self-force. If both parts 
are computed appropriately, the difference must become 
$R$-part, which is regular at least near the location of the particle. 
Therefore, the series 
\begin{equation}
 \sum_{\ell=0}^\infty 
\left(F_{full,\ell}^{\alpha}(\tau,R)
   -F_{S,\ell}^{\alpha}(\tau,R)\right)P_\ell (\cos X), 
\end{equation}
must be uniformly convergent. Thus we can take the coincidence limit, 
and we arrive at 
\begin{equation}
F_{R}^{\alpha}(\tau)
 =\sum_{\ell=0}^\infty 
\left(F_{full,\ell}^{\alpha}(\tau,R=0)
   -F_{S,\ell}^{\alpha}(\tau,R=0)\right).
\end{equation}

The $S$-part 
and hence also {\it full}-part of the self-force in 
the harmonic gauge takes the form 
of \cite{Barack:1999wf,MNS,BMNOS,Barack:2003mh}
\begin{equation}
F_{\ell}^\alpha(\tau,R=0)
 =A^\alpha \left(\ell+{1\over 2}\right) +
      B^\alpha +C^\alpha \left(\ell+{1\over 2}\right)^{-1} 
    +D^\alpha_\ell,
\end{equation}
The constants $A^\alpha$, $B^\alpha$, $C^\alpha$ and the 
residual part $D^\alpha:=\sum_{\ell=0}^\infty D^\alpha_\ell$ 
are called ``regularization parameters''. 
The choice of the residual for each $\ell$ mode, 
$D^{\alpha}_\ell$, can be changed rather freely by changing 
the behavior at a large separation angle, 
but the value of regularization 
parameter $D^\alpha$ is not affected. 
The first three regularization parameters 
$A^\alpha$, $B^\alpha$ and $C^\alpha$ must be common 
for both {\it full}- and $S$-parts since otherwise 
the $R$-part diverges. 
Hence, what we need to evaluate is the difference 
\begin{equation}
F_{R}^\alpha(\tau)
 = \sum_{\ell} \left(D^\alpha_{full,\ell}-D^\alpha_{S,\ell}\right). 
\end{equation}

The terminology ``mode-sum regularization'' or ``mode decomposition 
regularization'' is often used for the above prescription. 
However, we want to stress that 
the mode decomposition itself is just a useful tool 
to simplify the procedure of the point splitting regularization.

\subsection{gauge problem and use of intermediate gauge.}

The formal expression for the self-force is derived in harmonic 
gauge, and $S$-part is also given by using a local Hadamard 
expansion of the Green function in the harmonic gauge. 
However, when we calculate the full 
metric perturbations by using some metric reconstruction method 
from a master variable, the gauge that we can practically use 
is restricted to Regge-Wheeler-Zerilli gauge for Schwarzschild 
case or radiation gauge for Kerr case. 
Since the force is a gauge dependent quantity, 
regularization parameters can be different in different gauges. 
Hence, simple subtraction of $S$-part for harmonic gauge does not work. 

Here we present the basic idea of the intermediate 
gauge method~\cite{Barack:gauge} in a manner slightly modified 
from the original. 
We associate subscripts, $full, R, S$ to denote 
full perturbations, $R$-part and $S$-part, respectively. 
In any gauge $G$ we can define the $R$-part as 
\begin{equation}
 {\bf h}^{(G)}_R:={\bf h}^{(G)}_{full}-{\bf h}^{(G)}_S.  
\end{equation}
We consider a gauge transformation from the harmonic gauge to 
the gauge $G$:
\begin{equation}
 {\bf h}^{(G)}={\bf h}^{(H)}+\nabla \xi^{(H\to G)}[{\bf h}^{(H)}],
\end{equation}
where $(H)$ represents harmonic gauge, and $\nabla\xi$ denotes 
the change of metric 
generated by an infinitesimal coordinate transformation 
$x^\mu\to x^\mu- \xi^\mu$. 
$\xi^{(H\to G)}[{\bf h}^{(H)}]$ is the generator that 
transforms the metric perturbation in harmonic gauge 
${\bf h}^{(H)}$ into that in the gauge $G$. 
Then $R$-part of metric perturbations in the harmonic gauge can 
be rewritten as 
\begin{equation}
  {\bf h}^{(H)}_R={\bf h}^{(G)}_{full}-{\bf h}^{(G')}_S
   -\nabla\xi_R^{(H\to G)}, 
\label{hH}
\end{equation}
where 
\begin{equation}
 {\bf h}^{(G')}_S={\bf h}^{(H)}_S+\nabla \xi_S^{(H\to G)},  
\end{equation}
and 
\begin{equation}
\xi_R^{(H\to G)}
=\xi^{(H\to G)}[{\bf h}^{(H)}_{full}] -\xi_S^{(H\to G)}. 
\label{xixi}
\end{equation}
Here $\xi_S^{(H\to G)}$ is not specified yet, and 
it is not necessarily equal to $\xi^{(H\to G)}[{\bf h}_S^{(H)}]$. 
This is the reason why we used the notation ${\bf h}_S^{(G')}$ instead of 
${\bf h}_S^{(G)}$. 

The idea is to drop the last gauge term 
$\nabla\xi^{(H\to G)}[{\bf h}^{(H)}_R]$ in~(\ref{hH}). 
Namely, the force is computed from the regularized metric 
\begin{equation}
  {\bf h}^{(int)}_R={\bf h}^{(G)}_{full}-{\bf h}^{(G')}_S.
\end{equation}
A trajectory in this intermediate gauge is related to that 
in {\it the harmonic gauge} via the gauge transformation specified by 
$\xi_R^{(H\to G)}$. 
For the intermediate gauge to be useful, 
it should satisfy the following two conditions.
\vspace*{3mm}
\begin{enumerate}
\item We need to choose $\xi_S^{(H\to G)}$ so that  
${\bf h}_S^{(G')}$ completely 
cancels the singular part in ${\bf h}_{full}^{(G)}$ 
in the coincidence limit.
\item There is no secular growth 
in the generator of the gauge transformation $\xi_R^{(H\to G)}$. 
\end{enumerate}
\vspace*{3mm}
The condition (1) is the requirement that $\xi_S^{(H\to G)}$ is a 
good approximation of the true $S$-part gauge transformation, 
$\xi^{(H\to G)}[{\bf h}^{(H)}_S]$. 
If the difference between 
$\xi_S^{(H\to G)}$ and $\xi^{(H\to G)}[{\bf h}^{(H)}_S]$ 
is regular and finite, we have 
$\xi_R^{(H\to G)}= \xi^{(H\to G)}[{\bf h}_{R}^{(H)}]$ 
except for regular term. Hence, the coincidence limit of 
$\xi_R^{(H\to G)}$ is guaranteed to be finite. 
To obtain a sufficiently good approximation 
of $\xi^{(H\to G)}[{\bf h}_{S}^{(H)}]$,  
we just need to know ${\bf h}^{(H)}_S$ as a local expansion near the 
trajectory, and that is the best we can do.  

Here it is very convenient if 
the gauge $G$ satisfies 
a property that the gauge parameters to transform 
metric perturbations in another gauge to the specified gauge $G$ 
are determined without temporal or radial integration. 
This means that the equations to determine the gauge parameters 
are solved locally on a sphere. Then 
transformation into $G$-gauge is fixed unambiguously 
just by looking at the perturbations in the vicinity of a sphere.  
We call such a gauge as 
a {\it Gauge Operationally Deterministic on a Sphere} (GODS). 
Regge-Wheeler gauge is GODS\cite{Sanjay}. 
Radiation gauge in Kerr case does not seem to be GODS in 
its original form, 
but there seems to be a modification which allows it 
to transform into GODS in the sense of expansion with 
respect to Kerr parameter $a$\footnote{
We would like to come back to this point in future publication. 
}. 

In understanding the true difficulty about the gravitational 
radiation reaction, it is very important 
to realize that the condition (2) is really necessary. 
First of all, if the condition (2) is not required, 
we can add any finite gauge transformation. 
As a result, the trajectory in the coordinate representation 
can be arbitrarily changed, 
which is quite unsatisfactory situation. 
If the amplitude of gauge parameters is large, the coordinate 
values of the particle in $H$ and $G$ gauges are quite different. 
Even in that case, although it is quite 
counter-intuitive, one can cancel the divergent pieces 
in metric perturbations as far as linear perturbation is 
concerned, in which the gauge transformation does not couple 
with metric perturbations. However, once we consider the 
second order perturbation, large gauge parameters will cause 
disaster to the perturbative expansion. 

Now we come back to the issue how to guarantee the condition (2). 
When the gauge $G$ is GODS, 
we can define $\xi_S^{(H\to G)}$ 
in terms of local quantities without including any integration 
over $t$ or $r$.  
With such a choice of $\xi_S^{(H\to G)}$, 
we do not have to worry about the condition (2) for $S$-part  
since there is no secular growth in $\xi_S^{(H\to G)}$. 
What we need to guarantee to satisfy the condition (2) 
is the absence of secular growth in 
$\xi^{(H\to G)}[{\bf h}^{(H)}_{full}]$. 

We explain one practical 
scheme to guarantee the absence of secular growth in 
$\xi^{(H\to G)}[{\bf h}^{(H)}_{full}]$\footnote{
There is another way given by Amos Ori during post-Capra discussion 
meeting (2003 Kyoto), 
which does not rely on the presence of GODS.  
We give a brief explanation of his argument here
although it might be inaccurate. 
When the source can be decomposed into Fourier mode in 
time direction, both ${\bf h}^{(H)}_{full}$ and 
${\bf h}^{(G)}_{full}$ corresponding to a partial wave will be periodic. 
Namely, we assume 
${\bf h}_{full}(t,{\bf x})={\bf h}_{full}(t+T,{\bf x})$. 
Then, the gauge transformation connecting between two such 
metrics $ \nabla_{(\mu}\xi_{\nu)}$ satisfies the 
same periodicity. Hence, we have 
$
[\xi_{\mu;\nu} + \xi_{\nu;\mu}](t,{\bf x})=
[\xi_{\mu;\nu} + \xi_{\nu;\mu}](t+T,{\bf x}).  
$
This implies that $\xi_{\mu}(t+T,{\bf x})-\xi_\mu(t,{\bf x})$ 
satisfies the Killing equation. Therefore, 
$\xi_{\mu}={\rm (periodic~piece)} +
t K_\mu$, where $K_\mu$ is a Killing vector. 
For black hole background the variation of $K_\mu$ is limited, and hence for 
any choice of non-vanishing $K_\mu$, 
the gauge transformation $\nabla_{(\mu} (t K_{\nu)})$
does not vanish at $r^*\to \pm \infty$.  
As far as the ${\bf h}^{(G)}_{full}$ is guaranteed to go to zero 
at infinity or on the horizon, we find $K_\mu=0$.
Hence, the {\it full}-part of the gauge transformation 
does not have any secular growth.}. 
We first recall that ${\bf h}^{(H)}_{full}$ must stay finite by assumption. 
If this assumption does not hold, 
the whole formulation of self-force based on harmonic gauge breaks down, 
and hence we lose the whole foundation. 
For GODS, $\xi^{(H\to G)}[{\bf h}^{(H)}_{full}]$ is 
guaranteed to stay finite since so ${\bf h}^{(H)}_{full}$ is.

\section{Other topics}
Here we briefly mention a few topics which 
we have not yet discussed at all.

\subsection{analytic approach to the self-force}
There are a few calculation of the instantaneous self-force using 
numerical approach for a limited class of orbits\cite{Barack:2000zq}.
If we do not use the fully numerical approach, 
the {\it full}-part of the self-force is given in 
the form of Fourier expansion in time. On the other 
hand, the $S$-part is given in the local expansion, 
and hence is expressed in the time domain. 
To perform the subtraction, we need to transform 
the {\it full}-part (or the $S$-part) into the expression 
into the time domain (or into the frequency domain). 
As was mentioned earlier, a systematic method to solve the 
linear perturbation equation analytically is 
already known\cite{ManoTak,ManoTak2}. 
Taking advantage of it, 
we proposed a method to perform this transformation 
between different domains analytically\cite{analytic}.

\subsection{$\ell=0$ and $\ell=1$ modes}

There was a debate on how to treat $\ell=0$ and $\ell=1$ 
modes~\cite{Sago02,Detweiler03}. 
Here we give a simple-minded understanding of this complicated issue 
from the view-point of the intermediate gauge approach, 
although deeper understanding might indeed become indispensable 
when we consider the second order perturbation. 

The treatment of these lower lying modes in black hole perturbation 
theory is different from the other modes. 
The Regge-Wheeler-Zerilli 
formalism can handle these modes separately, 
but their equations are not hyperbolic. 
Therefore the retarded boundary conditions do not fix 
the boundary conditions for these modes. 
In the Teukolsky formalism, these modes are absent from the 
beginning. 

Those $\ell=0$ and 1 modes are composed of physical part and gauge part. 
In the intermediate gauge approach, no debate can 
arise in the gauge part. As long as it does not secularly increase, 
we do not care about it at all. 
The problem arises only in the point how to determine the 
physical part. There exists some information which is not 
encoded in the master variable, which is what we call 
$\ell=0$ and $1$ modes. 

The particle motion defines a three dimensional tube in 
four dimensional spacetime specified by 
the trajectory $(t,r)=(t_z(\tau),r_z(\tau))$. 
This tube divide the background spacetime into two 
pieces: one containing the infinity $i^0$ and the other containing 
black hole horizon. 

In both inner and outer regions, 
the metric perturbation is a homogeneous 
solution of Einstein equations. It is given by 
the part reconstructed from the master variables with $\ell\geq 2$ 
with an additional piece which does not affect the master variables 
with $\ell\geq 2$. Possibly such additional perturbations are 
those which are given by a change of the parameters contained 
in the background metric:
\begin{equation}
 \delta h_{\mu\nu}={\partial g^{BH}_{\mu\nu}\over \partial M}\delta M+
                   {\partial g^{BH}_{\mu\nu}\over \partial a}\delta a, 
\end{equation}
where $M$ and $a$ are the mass and Kerr parameter of the 
central black hole, respectively. 
For vanishing of the master variable, 
no other perturbations are possible. 
The Schwarzschild case should be considered as 
a special case of Kerr. If we consider an orbit with inclination, 
one may think that these two parameters are insufficient because 
we do not have parameters corresponding to the rotation 
in $x$ and $y$ axes. 
However, these rotations other than that in the $z$-direction can be 
absorbed by a gauge transformation, i.e., by global rotation of 
angular coordinates. 

If we do not care about small error of $O(\mu)$ in the estimate of the 
mass and the angular momentum of the central black hole, $\delta M$ and 
$\delta a$ are unimportant. They are renormalized in the definition 
of the mass and the angular momentum of the central black hole. 
Hence, this issue about $\ell=0$ and 1 is not an issue of debate 
at the lowest order in $\mu$. It becomes an issue only when 
we discuss it in connection with the second order perturbation.

\section{Summary}
In this paper, we gave a brief review of the recent 
development in the study of gravitational radiation
reaction problem in the context of generating gravitation 
wave templates, with some new insights. 

First, we reported the adiabatic approximation to the 
radiation reaction, in which long time average is assumed 
in evaluating the change rates of the ``constants of motion''. 
The radiation reaction to the Carter constant had been 
a long-standing issue, but now we are ready to compute 
it in the adiabatic approximation. We explained the 
formulation to evaluate the change rate of the Carter 
constant.
We have presented 
a method to integrate the evolution of orbits for a long 
period taking into account the radiation reaction. 
We have shown that the errors in the phases of the orbits
obtained by using the adiabatic approximation are 
$O(\mu\Delta\lambda,
\mu^2\Delta\lambda^2)$ or higher, where $\mu$ is the mass 
of the small compact star orbiting the central black hole and 
$\Delta\lambda$ is the time duration of integrating the 
orbit. In this order counting, $\Delta\lambda$ is 
supposed to be large, but $\mu\Delta\lambda$ is small. 
Typically, the $\Delta\lambda\propto \mu^{-1}$ since 
the evolution due to radiation reaction is slower for 
a smaller mass. 
The phase errors caused by ignorance of 
the instantaneous leading order self-force 
is $O(\mu\Delta\lambda)$. But when those errors become large, 
the contribution from the second order self-force
of $O(\mu^2\Delta\lambda^2)$ 
becomes comparable. 
However, there is a possibility that 
further study on the leading order self-force alone, 
beyond the level of adiabatic approximation, 
can improve the gravitational wave templates 
dramatically in some cases. 
As the orbit evolves adiabatically, the frequencies of 
oscillating part of the self-force change, and eventually 
one of them may cross zero.  
A rough order of magnitude estimate 
suggests that the corrections in phases 
due to this accidental appearance of zero frequency modes
can be as large as $O(\sqrt{\mu}\Delta\lambda)$.  
For more definite estimate of this effect we need to 
study the instantaneous self-force. 

Furthermore, once we start to detect gravitational waves from 
binary inspiral with an extreme mass ratio, the detailed 
comparison of the signal with the theoretical prediction 
will become possible. 
In such a situation, we will wish to have a theoretical tool 
which can predict the phase evolution with an accuracy 
less than $O(1)$. For this purpose, we need to understand 
the self-force not only at the leading order but also 
up to the second order, although probably some kind of 
averaged values will be sufficient for the second order 
self-force. 

Also as a step toward this goal, complete understanding about 
the first order self-force will be necessary. Recently, 
there have been a lot of developments also in this direction. 
We have presented here a biased summary about this issue, partly 
due to my understanding and also due to lack of space.  
So far, the lowest order self-force has not been 
discussed extensively as a step toward the second 
order self-force\cite{Rosenthal05}.
In order to reflect the stored knowledge about the self-force 
to the improvement of the theoretical prediction of gravitational 
waveforms, we need to develop a formalism 
to evaluate the second order self-force.

\section*{acknowledgement}
The author thanks Hiroyuki Nakano, Norichika Sago, 
Wataru Hikida and Sanjay Jhingan 
for their valuable comments and careful reading of the 
manuscript. 
The discussion during the post-YKIS program 
``Gravity and Cosmology'' was also very meaningful to complete 
this article. 
This work is supported in part 
by Grant-in-Aid for Scientific Research, Nos.
14047212, 14047214 and 12640269, 
and by that for the 21st Century COE
"Center for Diversity and Universality in Physics" at 
Kyoto university, both from the Ministry of
Education, Culture, Sports, Science and Technology of Japan.

\end{document}